\documentclass[preprint,showpacs,preprintnumbers,amsmath,amssymb,
superscriptaddress]{revtex4}

\usepackage{graphicx}
\usepackage{dcolumn}
\usepackage{bm}

\begin{document}

\title{Reply to the "Comment on 'Precise half-life values for two-neutrino double beta decay"}

\author{A.S.~Barabash} \email{barabash@itep.ru}
\affiliation{Institute of Theoretical and Experimental Physics, B.\
Cheremushkinskaya 25, 117218 Moscow, Russian Federation}

\date{\today}


\pacs{23.40.-s, 14.60.Pq}


\maketitle

Addressed here is Comment \cite{PRI10} which suggests that in work \cite{BAR10} an incorrect (underestimated) 
value for a half-life of $^{128}$Te is obtained and that \cite{PRI10,PRI10a} yield a more accurate 
estimation of it. I will speak to a number of disputable remarks in \cite{PRI10}. 

1. Comment (2-nd paragraph): "In the analysis, author goes through extensive selection,
 removal of discrepant data sets and adjustments procedures for $^{128}$Te geochemical data sets 
using his previous work \cite{BAR00} on time variation of weak interaction constant as an explanation."

B. Pritychenko has probably not understood the estimation of $T_{1/2}$($^{128}$Te) in this work \cite{BAR10}.  
It is not an averaging procedure. To minimize the possible uncertainties in the $T_{1/2}$($^{128}$Te) 
the value from direct counting experiments 
$T_{1/2}$($^{130}$Te) = $(6.8^{+1.2}_{-1.1})\times 10^{20}$ y \cite{BAR10} and  
the well-known ratio 
of $T_{1/2}$($^{128}$Te) / $T_{1/2}$ ($^{130}$Te) = $(2.84 \pm 0.09) \times 10^3$ \cite{BER93} 
were used. Multiplying 
one value by the ratio yields the value $T_{1/2}$ ($^{128}$Te) $= (1.9 \pm 0.4) \times 10^{24}$ y. 
I consider this to be 
the most currently reliable and accurate estimation for $T_{1/2}$($^{128}$Te).

2. Comment (3-d paragraph): "Additionally, one cannot reject 
$^{128}$Te $T_{1/2}$ value from the Washington University group but still use the 
$^{128}$Te/$^{130}$Te ratio from the same group."
 
This remark is caused by misunderstanding of how the estimation of $T_{1/2}$($^{128}$Te) has been made 
as introduced above. Again, to estimate the value $T_{1/2}$($^{128}$Te) the well-known ratio $T_{1/2}$ ($^{128}$Te)/$T_{1/2}$ ($^{130}$Te) from work \cite{BER93} was used. In fact, it was argued 
in \cite{BAR10} and in many previous papers \cite{BER93,TAK96, MAN91,MES08} that with geochemical 
experiments one derives a more reliable results with a ratio rather than with half-life values 
for the individual isotopes. The problem with geochemical experiments is to establish correctly 
the gas-retention age. If, for example, there is a leak of xenon during the life of the mineral 
then one can obtain an overestimated value for the half-life of $^{130}$Te or $^{128}$Te. 
However this leak will not change the ratio because it will be the same leak for $^{130}$Xe 
and $^{128}$Xe (daughter nuclei of $^{130}$Te and $^{128}$Te). So, conveniently one does not 
need a precise age of the mineral (gas-retention age). Consequently, my estimation did not 
use separate results of geochemical experiments for $^{128}$Te. 

The results of different groups are not in agreement. I point out that 
$T_{1/2}$ ($^{128}$Te) $= (2.2 \pm 0.3) \times 10^{24}$ y was obtained in \cite{TAK96}, 
but $T_{1/2}$ ($^{128}$Te) $= (7.7 \pm 0.4) \times 10^{24}$ y was established in \cite{BER93}. 
So, the difference is more than $10 \sigma$. This means there is some problem with the 
data and following the recommendations of PDG, "we may choose not to use the average at 
all" \cite{PDG00}. Fortunately, the stable value of  $T_{1/2}$ ($^{128}$Te)/$T_{1/2}$ 
($^{130}$Te) is known \cite{BER93}, and recently $T_{1/2}$($^{130}$Te) was accurately 
measured with the NEMO-3 detector \cite{BAR10a}. This yields a  precise value for 
$T_{1/2}$($^{128}$Te) presented in  \cite{BAR10}.

3. Comment (4-th paragraph): "NNDC $^{128}$Te half-life value 
$(3.2 \pm 2.0) \times 10^{24}$ y is almost twice as large than of Barabash 
$(1.9 \pm 0.4) \times 10^{24}$ y...".
 
In fact, one can see that both values are in agreement taking into account the very 
large error in the NNDC case. The large error in the NNDC case is because the experimental 
results for $^{128}$Te are not in agreement with each other and these results 
were used in \cite{PRI10a} to obtain an average value. Generally speaking, this procedure 
is not correct (see remark to comment 2). 

4. Comment (4-th paragraph, equation (1)).
Using his own estimation for $^{128}$Te and $^{130}$Te and the idea that Nuclear Matrix 
Elements (NME) are the same for these nuclei, B. Pritychenko obtained the following ratio 
(1):
      
$T_{1/2}^{2\nu}$($^{128}$Te)/ $T_{1/2}^{2\nu}$($^{130}$Te) $ \approx$ $5.7 \times 10^3$ $\sim$ ($E_{130}/E_{128})^8$                                         (1)

This is correct if the difference in $T_{1/2}$ is attributed to a difference in the 
$2\nu$ transition energy only. But, in fact, there is very precise experimental data for this ratio:  $T_{1/2}^{2\nu}$($^{128}$Te)/ $T_{1/2}^{2\nu}$($^{130}$Te) $=  (2.84 \pm 0.09) \times 10^3$ \cite{BER93}. 
This indicates that the estimation (1) is non-correct and NMEs are not the same for these 
nuclei. B. Pontecorvo, many years ago (1968), made this assumption 
[NME($^{128}$Te) = NME($^{130}$Te)] \cite{PON68} and at that time it was quite a fruitful idea. 
Qualitatively, this assumption is correct even now because the difference is on the 
level $ \sim 50\%$. Since that time, progress in experiments \cite{BER93} and theory 
\cite{CAU99,SUH98,SIN07} indicate the "equality" is in contradiction with both experiment and theory. 

5. Comment (5-th paragraph): "Deviation from the nuclear structure evaluation policies in
 the work \cite{BAR10} produced underestimated $T_{1/2}$ value for $^{128}$Te \cite{BAR00} 
and distorted tellurium ratio for evaluated $T_{1/2}$."

   Again, this remark is from a misunderstanding of how the estimation of the $T_{1/2}$($^{128}$Te) was made as discussed above.  Concerning "...and distorted tellurium ratio for evaluated $T_{1/2}$", the ratio is fixed in the experiment \cite{BER93}. In addition, there is no theoretical argument for NME($^{128}$Te)=NME($^{130}$Te). The equality is not supported by the modern Shell Model \cite{CAU99}, QRPA \cite{SUH98} and PHFB calculations \cite{SIN07} which predict a difference between NME($^{128}$Te) and NME($^{130}$Te).  

In conclusion, this should clarify the criticisms in work \cite{PRI10} which appear to have come from a misunderstanding of the analysis in work \cite{BAR10}. I stand by my conclusions as presented \cite{BAR10}.

\section*{Acknowledgements}

I am very thankful to Prof. S. Sutton for his useful remarks. 
A portion of this work was supported by grant from RFBR 
(09-02-92676).
This work was also supported by the Russian Federal Agency for Atomic Energy.






\bibliographystyle{aipproc}   

\bibliography{sample}

\IfFileExists{\jobname.bbl}{}
 {\typeout{}
  \typeout{******************************************}
  \typeout{** Please run "bibtex \jobname" to optain}
  \typeout{** the bibliography and then re-run LaTeX}
  \typeout{** twice to fix the references!}
  \typeout{******************************************}
  \typeout{}
 }

\end{document}